\RequirePackage{fix-cm}

\documentclass[twocolumn,epjc3]{svjour3}  
\smartqed  
\RequirePackage{graphicx}
\usepackage{latexsym}
\usepackage{hyperref}
\usepackage{dcolumn}
\usepackage{bm}
\usepackage[export]{adjustbox}
\usepackage{siunitx}
\usepackage[square,comma,numbers,compress]{natbib}
\usepackage[utf8]{inputenc}
\usepackage{tabularx}
\usepackage{amsmath}
\journalname{Eur. Phys. J. C}
\begin{document}

\title{{Low-Energy Solar Neutrino Detection Utilizing Advanced Germanium Detectors}
}
\author{S. Bhattarai\thanksref{USD}
         \and
         D.-M. Mei\thanksref{e1,USD}
         \and
         M.-S. Raut\thanksref{USD}
          }
\thankstext{e1}{dongming.mei@usd.edu}

\thankstext{t1}{ This work was supported in part by NSF OISE 1743790, DOE grant DE-FG02-10ER46709, DE-SC0004768, the Office of Research at the University of South Dakota and a research center supported by the State of South Dakota.}


\institute{Department of Physics, University of South Dakota, 414 E Clark St, Vermillion, South Dakota 57069  \label{USD}
}

\date{Received: date / Accepted: date}

\maketitle

\begin{abstract}
We explore the possibility to use advanced germanium (Ge) detectors as a low-energy solar neutrino observatory by means of neutrino-nucleus elastic scattering. A Ge detector utilizing internal charge amplification for the charge carriers created by the ionization of impurities is a novel technology with experimental sensitivity for detecting low-energy solar neutrinos. Ge internal charge amplification (GeICA) will amplify the charge carriers induced by neutrino interacting with Ge atoms through emission of phonons. It is those phonons that will create charge carriers through the ionization of impurities to achieve an extremely low energy threshold of $\sim$0.01 eV.  We demonstrate the phonon absorption, excitation, and ionization probability of impurities in a Ge detector with impurity levels of 3$\times$10$^{10}$ cm$^{-3}$, 9$\times$10$^{10}$ cm$^{-3}$, and 2$\times$10$^{11}$ cm$^{-3}$. We present the sensitivity of such a Ge experiment for detecting solar neutrinos in the low-energy region. We show that, if GeICA technology becomes available, then a new opportunity arises to observe $pp$ and $^{7}$Be solar neutrinos. Such a novel detector with only 1 kg of high-purity Ge will give $\sim$ 10 events per year for $pp$ neutrinos and $\sim$ 5 events per year for $^{7}$Be neutrinos with a detection energy threshold of 0.01 eV. 
\end{abstract}

\section{Introduction}
\label{intro}
\begin{figure}
    
     \includegraphics[width=0.5\textwidth]{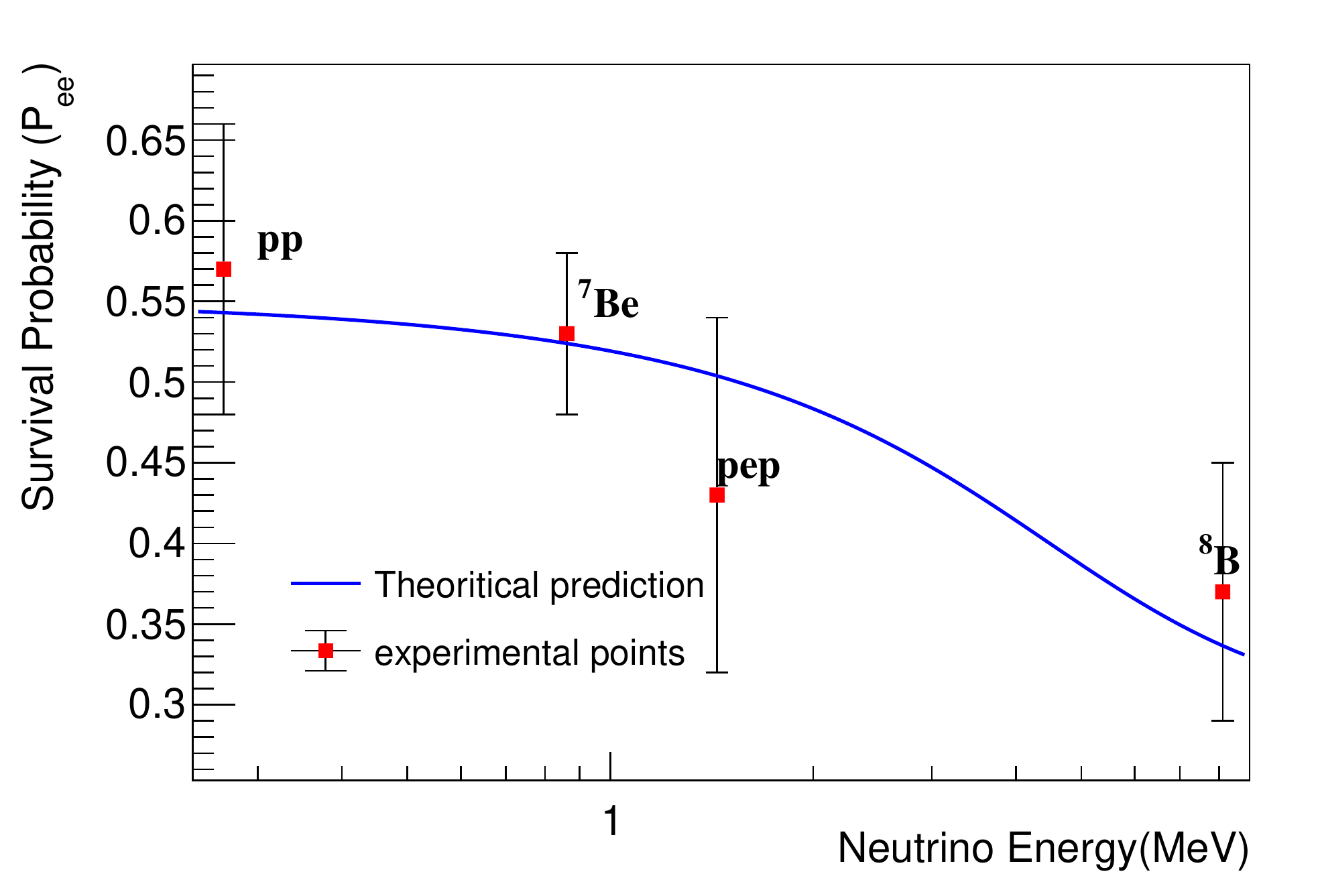}
    \caption{Survival probability of solar neutrinos from $pp$ chain}
    \label{fig:my_label1}
    \end{figure}

Solar neutrinos are the neutrinos produced in the core of the sun through nuclear fusion reactions. The study of solar neutrinos provides information about the fundamental properties of neutrinos and allows us to understand how the sun works. There are various nuclear fusion reactions occurring in the sun that produce solar neutrinos of different fluxes and their respective maximum energies~\cite{papoulias2018novel}. The solar neutrinos emitted in several steps of the proton-proton (P-P) cycle are $pp$, $pep$, $^7Be$, $^8B$ and $hep$ neutrinos. The solar neutrinos produced through the Carbon-Nitrogen-Oxygen (CNO) cycle are mainly ${}^{13}N$, ${}^{15}O$ and ${}^{17}F$ neutrinos~\cite{bahcall1998uncertain,brun1998standard}. The energy range of these neutrinos varies from a few keV to a few MeV. Detecting these neutrinos by placing a detector in a underground laboratory has always been a challenge. A suitable energy threshold for detecting each of these neutrinos is difficult to achieve. Numerous collaborations have used different detector materials and techniques to study solar neutrinos. Homestake~\cite{davis1994review}, Super-Kamaiokande~\cite{fukuda2001solar}, SNO~\cite{harrison2002tri}, BOREXINO~\cite{arpesella2008direct}, GALLEX/GNO~\cite{altmann2005complete}, SAGE~\cite{abdurashitov1994results}, LENS~\cite{cribier2000lens}, are some of the experiments that studied solar neutrinos with the detection energy threshold greater than 233 keV. By detecting the flux of neutrinos from the $8^B$ reaction in the sun, SNO was able to completely demonstrate neutrino flavor transition using charge current (cc) and neutral (nc) current reactions~\cite{harrison2002tri}. The comparison of reaction rate 
between cc and nc reactions help in the measurement of total electron neutrino ($\nu_e$) flux and the total flux independent of the flavor ($\nu_e,\nu_\tau,\nu_\mu$). A strong suppression of electron neutrinos was observed relative to that expected in the standard solar model (SSM)~\cite{guenther1992standard}. This indicated that the electron neutrinos from the sun are changing to other flavor of neutrinos by neutrino flavor transition. The most accurate measurement of neutrino survival probability to date is observed by Borexino~\cite{agostini2018comprehensive}. They used elastic neutrino-electron scattering to experimentally calculate the values of survival probability~\cite{agostini2018comprehensive} for four electron neutrinos from the $pp$ chain  which is shown in Figure~\ref{fig:my_label1}. There is a significant amount of difference between the theoretical~\cite{wolfenstein1978neutrino,wolfenstein1979neutrino,bahcall200610,esteban2017updated} and the experimental values as depicted in Figure~\ref{fig:my_label1}.

Over 99 percent of the total solar neutrinos produced from $pp$ cycles are believed to be $pp$ neutrinos~\cite{papoulias2018novel}. They exhibit a continuous spectrum with an end-point energy of 423 keV, which makes it difficult to be detected using liquid scintillation detectors. This is because the energy deposited through elastic neutrino-nucleus scattering falls below the detection threshold and the energy induced by elastic neutrino-electron scattering is often contaminated by backgrounds. Therefore, the measured event rate from $pp$ neutrinos can be largely uncertain as shown by Borexino~\cite{arpesella2008direct}. 

For a solid state detector, such as a Ge detector, both elastic 
neutrino-nucleus scattering and elastic neutrino-electron scattering can be detected. However, since the detectors are much smaller in size compared to liquid scintillation detectors, the energy deposited through elastic neutrino-electron scattering is often immersed in backgrounds. Nevertheless, the event rate grows exponentially as a function of nuclear recoil energy through elastic scattering off the nucleus. It is expected that the event rate will significantly surpass the background in the low energy range of nuclear recoils. This is because the expected nuclear recoil events (signal) grows exponentially while electric recoil events (background) remains flat in the region of interest.  Thus, elastic neutrino-nucleus scattering represents a viable tool to measure $pp$ neutrinos in a Ge detector. The maximum nuclear recoil energy produced by $pp$ neutrinos via elastic neutrino-nucleus scattering in a Ge detector is $\sim$5.2 eV. Hence, to detect the $pp$ solar neutrinos, an extremely low-energy threshold detector is needed. CDMS~\cite{cdms2008search}, SuperCDMS~\cite{agnese2014search} and EDELWEISS~\cite{armengaud2013background} have demonstrated that the energy threshold of $\sim$50 eV to $\sim$ 100 eV can be achieved in Ge detectors through detecting phonons. In 2018, SuperCDMS reported  a 3 eV phonon energy resolution with a 0.93-gram Si detector when biased at 100 V~\cite{agnese2018first}.

Nevertheless, nuclear recoils induced by $pp$ neutrinos require detectors of threshold lower than 1 eV to have meaningful statistics. Therefore, because the current state-of-the-art Ge detectors cannot detect the $pp$ neutrinos through elastic neutrino-nucleus scattering, a new type of detector is required. A Ge internal charge amplification (GeICA) detector, that amplifies the charge carriers created by the ionization of impurities, is a novel technology with experimental sensitivity for detecting the low-energy solar neutrinos~\cite{mei2018direct}. With an extremely low-energy threshold ($\sim$0.01 eV), GeICA detectors can measure the $pp$ neutrinos flux through coherent elastic neutrino-nucleus scattering with good statistics and hence the current uncertainty in neutrino survival probability (Figure~\ref{fig:my_label1}) can be decreased. In this paper, we describe the GeICA detector technology for achieving a sensitivity in detecting low-energy $pp$ neutrinos. GeICA will amplify the charge carriers induced by $pp$ neutrinos interacting with Ge atoms through the emission of phonons\cite{mei2018direct}. It is those phonons that will create charge carriers through the ionization of impurities to achieve an extremely low energy threshold of $\sim$ 0.01 eV.

\begin{figure}
    \includegraphics[width=0.5\textwidth]{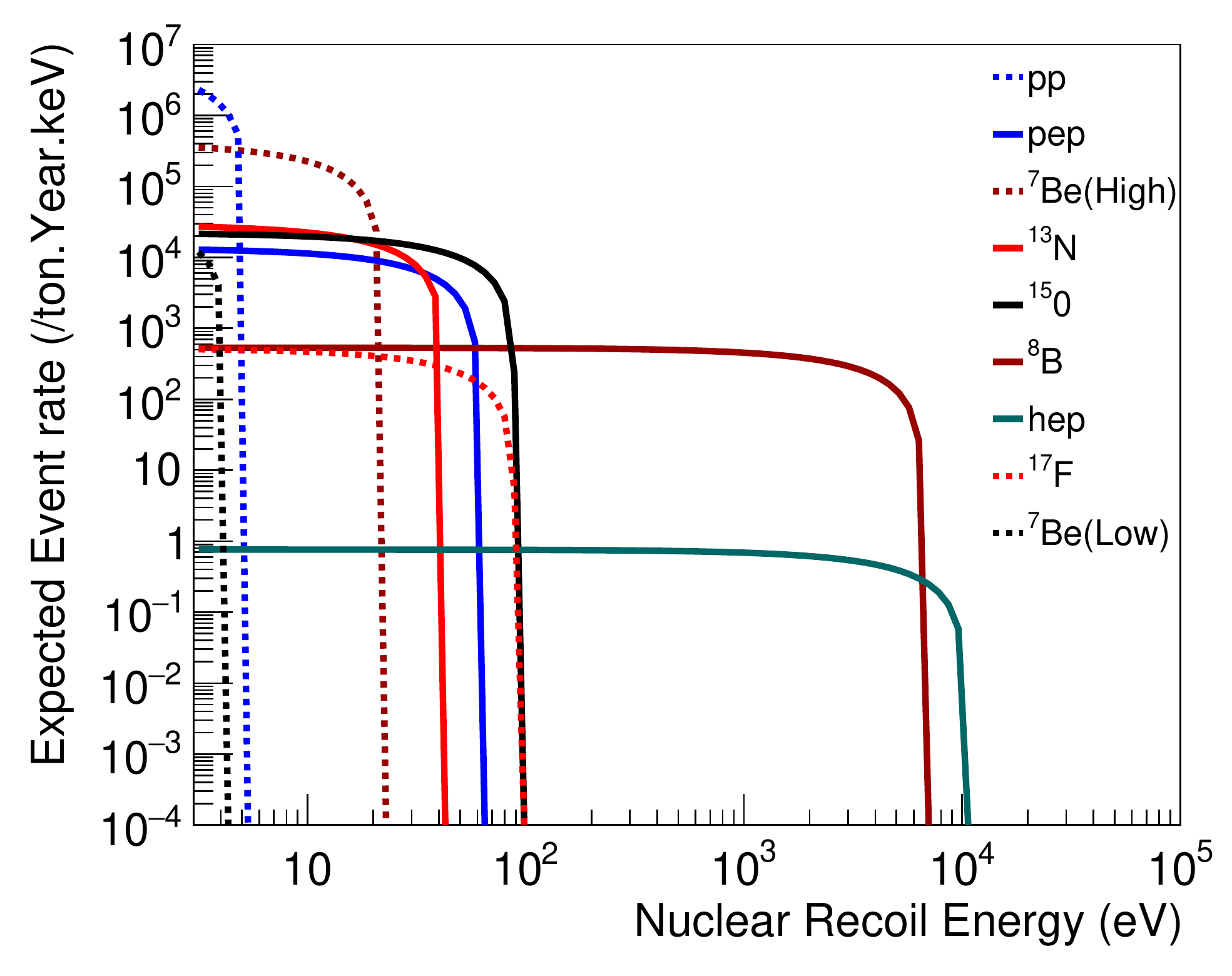}
    \caption{The expected event rate versus nuclear recoil energy produced by solar neutrinos in a Ge target. }
    \label{fig:my_label2}
\end{figure}

Coherent elastic neutrino-nucleus scattering has not been used for detecting $pp$ neutrinos because of the low amount of energy transferred to a nucleus during the interaction. However, utilizing internal charge amplification, the charge  carriers created by phonon excitation can be used to detect $pp$ neutrinos because of the extremely low energy threshold of the detector\cite{starostin2000germanium}. In addition, the size of the detector can be dramatically reduced. The event rate from coherent elastic neutrino-nucleus scattering is much higher than that of elastic neutrino-electron scattering. The differential neutrino-nucleus cross-section $d\sigma_{CNS}(E_\nu,E_{NR})/dE$ for a neutrino of energy $E_\nu$ (eV) is given by
\begin{equation}
    \frac{d\sigma_{CNS}(E_\nu,E_{NR})}{dE_{NR}}=\frac{{G_F}^2}{4\pi}{Q_w}^2m_N(1-\frac{m_NE_{NR}}{2{E_\nu}^2})F^2(T_R)
    \label{equation_1}
\end{equation} 
where $E_{NR}$ is nuclear recoil energy, $m_N$ is the mass of the target nucleus, $G_F$ is the Fermi Coupling constant, $Q_w$=$N-(1-4Sin^2\theta_w)Z$ where $N$ the number of neutrons and $Z$ is the number of protons and  $\theta_w$ the mixing angle. Here, the value of the form factor $F(T_R)$ is equal to 1~\cite{billard2015solar}.  Likewise, the differential event rate for a detector of mass $M$ and exposure time $T$ is given by \begin{equation}
    \frac{dE}{dE_{NR}}=N_T\times M \times T \times\frac{d\sigma_{CNS}(E_\nu,E_{NR})}{dE_{RN}}\times\frac{dN_\nu}{dE_\nu}
    \label{equation_2}
\end{equation}
where $N_T$ is the number of target nuclei per unit mass and $\frac{dN_\nu}{dE_\nu}$ is the differential neutrino flux. Figure~\ref{fig:my_label2} shows the expected event rate as a function of nuclear recoil energy induced by solar neutrinos in a Ge detector. Through coherent elastic neutrino-nucleus scattering, we can detect $pp$ neutrinos in different flavors ($\nu_e,\nu_\tau,\nu_\mu$) without considering the neutrino oscillation. This leads to the determination of total neutrino flux  in the detector. Similar to the SNO experiment, which determined the total $^8B$ neutrino flux and hence proved the neutrino flavor transition, measuring the total $pp$ neutrino flux will verify the standard solar model and the neutrino flavor transition at lower neutrino energy when combined with the global measurements for the solar neutrino survival probability, as shown in Figure~\ref{fig:my_label1}.  

\begin{table}
\centering
  \resizebox{\linewidth}{!}{
  \begin{tabular}{|c|c|c|}
 \hline
 Type &  Maximum nuclear recoil& Total Event $(/kg.year)$ \\ 
 \hline
  pp & 5.29 & $16.58$\\ 
 
  pep & 61.32 & $0.46$  \\
  hep & 10386.03& $0.001$ \\
  ${}^7Be_{low}$&4.27&$0.11$ \\
  ${}^7Be_{high}$&21.87&$4.98$\\
  ${}^8B$&6654&$1.86$\\
  ${}^{13}N$&42.58&$0.67$\\
  ${}^{15}O$&88.71&$1.09$\\
  ${}^{17}F$&89.53&$0.03$\\
 
 \hline
 \end{tabular}
 }
 \caption{Maximum nuclear recoil and total event rate integrated for solar neutrinos.}
  \label{tab:my_table1}

\end{table}
 
Below, we demonstrate how $pp$ neutrinos can be measured more accurately with the proposed Ge detector that uses phonons generated by neutrinos via coherent neutrino-nucleus elastic scattering to ionize impurities.
\section{\label{sec:level2}{The Working Principle of the Proposed Detector}}
Starostin et al.~\cite{starostin2000germanium} have proposed a detector that  can be used to amplify the signal generated by the  pp solar neutrinos. It is assumed to be made from a 1.0 kg  HPGe crystal. The net impurity concentration in the detector will be $(1-3)\times10^{10}/cm^3$. It  will be a multi strip planar Ge detector having a dimension of $9 \text{cm} \times 7 \text{cm} \times 3 \text{cm}$ with 15 anode strips fabricated using the photo-mask method each of width 20 $\SI{}{\micro\metre}$. The fiducial volume of the detector will be about 190 $cm^3$~\cite{mei2018direct,starostin2000germanium}.  The detector concept and its working principle were discussed in detail in our earlier publication~\cite{mei2018direct}. The main conclusions are:
\begin{enumerate}
\item{
After purifying Ge ingots to a level of $\sim$10$^{11}$/cm$^{3}$ by zone refining~\cite{yang2014investigation}, a single crystal can be grown at USD through the Czochralski method~\cite{czochralski1918neues}; During the crystal growth process, impurities can be further removed from the grown crystal down to a level of $\sim$10$^{10}$/cm$^{3}$ or below~\cite{wang2012development}. }
\item{It has been found that the remaining impurities are mainly  aluminium (Al), phosphorous (P), boron (B) and gallium (Ga) in the USD-grown crystals~\cite{czochralski1918neues}; The ionization energies of these impurities in Ge are in a level of $\sim$0.01 eV,  which is less than the longitudinal acoustic (LA) phonon (0.04 eV) and the transverse (TA) phonon (0.026 eV) generated by neutrinos via coherent neutrino-nucleus elastic scattering~\cite{wittmann2007miniaturization,mei2018direct};  Hence the phonons can certainly excite or ionize these impurities to produce charge carriers. }
\item{These charge carriers will then be drifted towards the electrical contacts. During the drifting process, these charge carriers will be accelerated by a high electric field to generate more charge carriers and hence, amplify the charge by a factor of $\sim$100 to $\sim$1000, depending  on  the  applied  electric  field.}
\item{The absorption probability $P$ of phonons in a given Ge detector can be estimated as
\begin{equation}
    P=1-exp(-d/\lambda)
    \label{equation_6}
\end{equation}
where $d$ is the average distance diffused before an anharmonic decay and $\lambda=\frac{1}{\sigma  \times N_A}$ is the mean free path of phonons with $N_A$ being the net impurity level in a given p-type detector and $\sigma$ is the cross-section of phonons absorbed by neutral impurities~\cite{mei2018direct}.}
\item{Similarly, the ionization or excitation probability  ($f(E_A)$) of a neutral acceptor state to be ionized is given by
\begin{equation}
    f(E_A)=1-\frac{1}{1+4e^{(E_A-E_F)/k_BT}}
    \label{equation_7}
\end{equation}
where $E_F$ is the Fermi energy level and $(E_F- E_V )$ = $k_BTln(N_V/ N_A )$ with $N_V = 2(2\pi m^*k_BT/h^2)^{3/2}$ being the effective states, $m^*$ is the eﬀective mass of a hole,$k_B$ is the Boltzmann constant~\cite{mei2018direct}. However, the recoil energy produced by $pp$ neutrinos can only produce a few charge carriers by exciting these impurities. Such a small signal would be immersed in the noise of the generic Ge detectors. If the charge carriers can be internally amplified to surpass the level of electronic noise, then such a small signal created by $pp$ neutrinos can be detected by a GeICA detector.}
 \end{enumerate}
 \section{\label{sec:level3}{Absorption Cross Section}}
A critical question related to the above detector working principle is the phonon absorption cross section. The scattering mechanism of phonons off the neutral impurities at low temperature regime is governed by the following equations
\begin{equation}
    D^X+\Delta E_X\rightarrow e^-+D^+;
    A^X+\Delta E_A\rightarrow h^++A^-
\end{equation}
where $D^X$ represents neutral donors, $\Delta E_D$ is the energy absorbed by neutral donors, $e^-$ is the charge carrier produced after ionization of impurity. Here, $\Delta E_D$ is the energy of incoming phonon. The absorption cross section is independent of the incoming particle but rather depends upon incoming particle's energy (frequency). The net impurities present in our crystal for this work is $2\times10^{10}/cm^3$ and the energy of phonons we have used ranges from 0.00325 eV to 0.026 eV. Using a direct analogy to photons having energy $\hbar\omega$ as quanta of excitation of the lattice vibration mode of angular frequency $\omega$, the angular frequency of phonons with energy from 0.00325 eV to 0.026 eV is in the range of 4.92$\times$10$^{12}$Hz to 3.94$\times$10$^{13}$Hz. According to Majumdar~\cite{majumdar1993microscale}, the cross section of scattering of lattice waves (phonons) off an impurity with radius $R$ is given by 
\begin{equation}
    \sigma=\pi R^2\chi^4/(\chi^4+1),
\end{equation}
where $\chi=\omega R/v$ is called the size parameter, $v$ is the group velocity of phonons assumed to be a constant, which is equal to the speed of sound in Ge, $\omega$ is the angular frequency of phonons. For calculating the effective radius of impurities in Ge,  we have used the effective mass approximation~\cite{skinner2019properties} with effective Bohr's radius ($R$) of impurities:
\begin{equation}
    R= (0.53\AA) \epsilon/(m^*/m)
\end{equation}
where $\epsilon$ is the dielectric constant of Ge which is equal to 16, $m^*$ is the hydrogenic effective mass~\cite{dubon1996electronic}. The value of $m^*/m$ is taken to be 0.21 in the case of holes in $P$-type Ge. Using these values we have calculated the values of cross-sections for various energies of phonons which is depicted in Table ~\ref{tab:my_label3}. From Table~\ref{tab:my_label3} we can infer that, for the range of the phonon energies we have used in this work, the absorption cross sections is almost a constant, which is $\sim$5$\times$10$^{-13}$cm$^{2}$. \\
\begin{table*}

    \begin{tabular}{|c|c|c|c|c|c|c|}
    \hline
        Phonon energy($eV$)& Frequency(Hz)& Effective radius($cm$)& Size parameter $\chi$&$\chi^4/(\chi^4+1)$&Cross-section($cm^2$)  \\
        \hline
         0.037&$5.61\times10^{13}$&$4.03\times10^-7$&$42.01$&1&$5.12\times10^{-13}$\\
         \hline
         0.026&$3.94\times10^{13}$&$4.03\times10^-7$&$29.52$&0.999&$5.12\times10^{-13}$\\
         \hline
         0.013&$1.51\times10^{13}$&$4.03\times10^-7$&$11.35$&0.999&$5.11\times10^{-13}$\\
         \hline
         0.0065&$9.87\times10^{12}$&$4.03\times10^-7$&7.38&0.999&$5.11\times10^{-13}$\\
         \hline
         0.00325&$4.93\times10^{12}$&$4.03\times10^-7$&3.69&0.994&$5.09\times10^{-13}$\\
         \hline
    \end{tabular}
    \caption{ Phonon-impurity cross-sections for various energies of phonons and their corresponding angular frequencies. }
    \label{tab:my_label3}
\end{table*}
  
\section{\label{sec:level4}{Projected Sensitivity}}
\begin{figure}
    
     \includegraphics[width=0.5\textwidth]{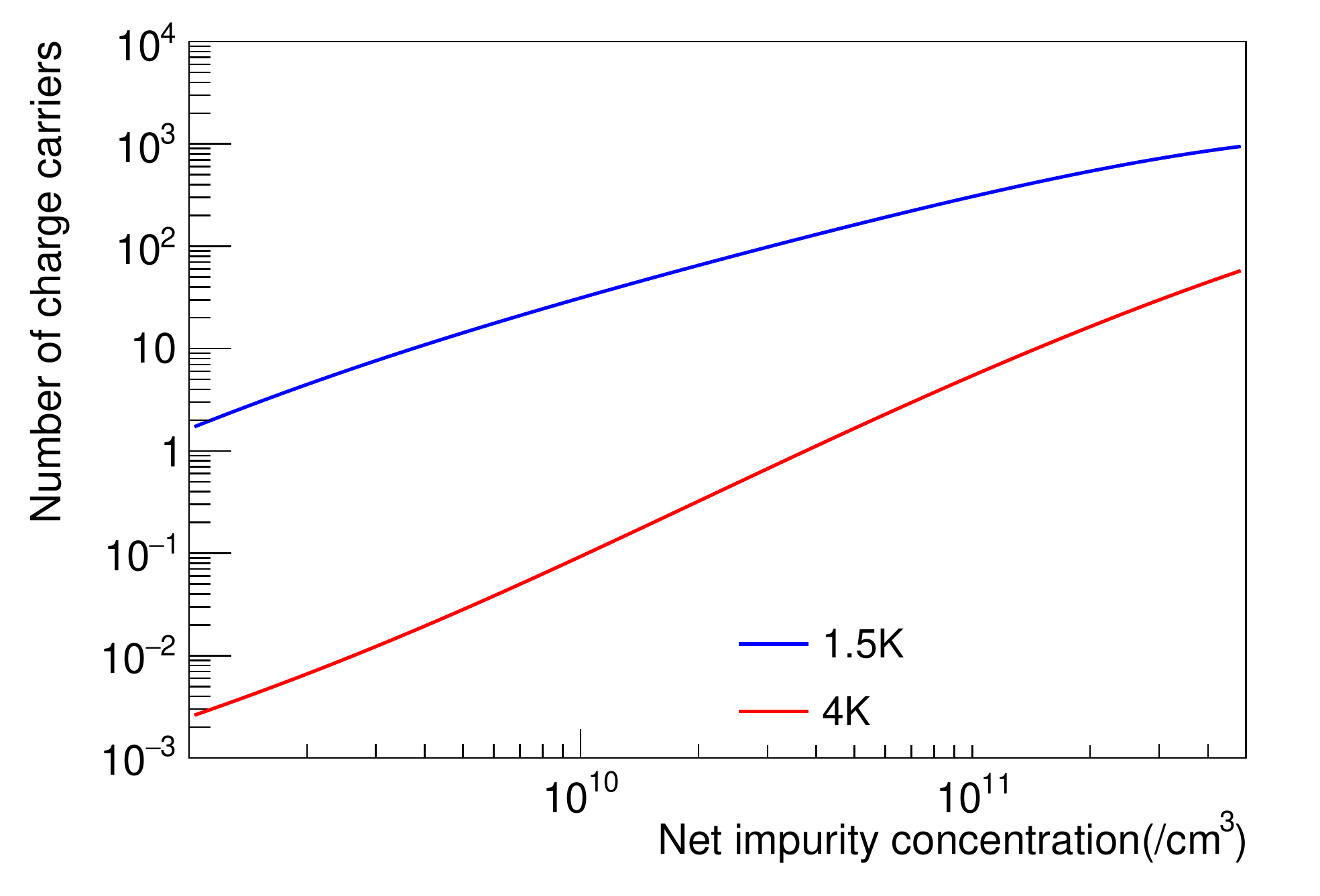}
    \caption{Number of charge carriers for different impurities in a Ge detector when the detector is operated at very low temperature of 1.5 K and 4 K.}
    \label{fig:my_label5}
\end{figure}

If we assume a Ge detector of 3 cm thickness with a total energy deposition of 2.0 eV, where each phonon has the energy of 0.026 eV, then the total number of parent phonons is $\sim$76. During the transport, each parent phonon undergoes anharmonic decay to generate two daughter phonons, each of which has energy equal to half of the parent phonon. We can estimate the total number of charge carriers $N_{carriers}$ using the formula below:
\begin{equation}
   N_{carriers}=\sum\limits_{i}n_ip_if(E_A)_i
   \label{equation_8}
\end{equation} 
where $n_i$ is the number of $i^{th}$ phonons, $p_i$ is the absorption probability of $i^{th}$ phonons given by equation~\ref{equation_6} and $f(E_A)_i$ is the ionization or excitation probability of $i^{th}$ phonons given by equation~\ref{equation_7}. Figure~\ref{fig:my_label5} shows the total number of charge carriers created by the ionization or excitation of impurities as a function of impurity level in a given Ge detector with energy threshold of 2.0 eV for two temperatures, 1.5 K and 4 K.
\begin{figure}
    
     \includegraphics[width=0.5\textwidth]{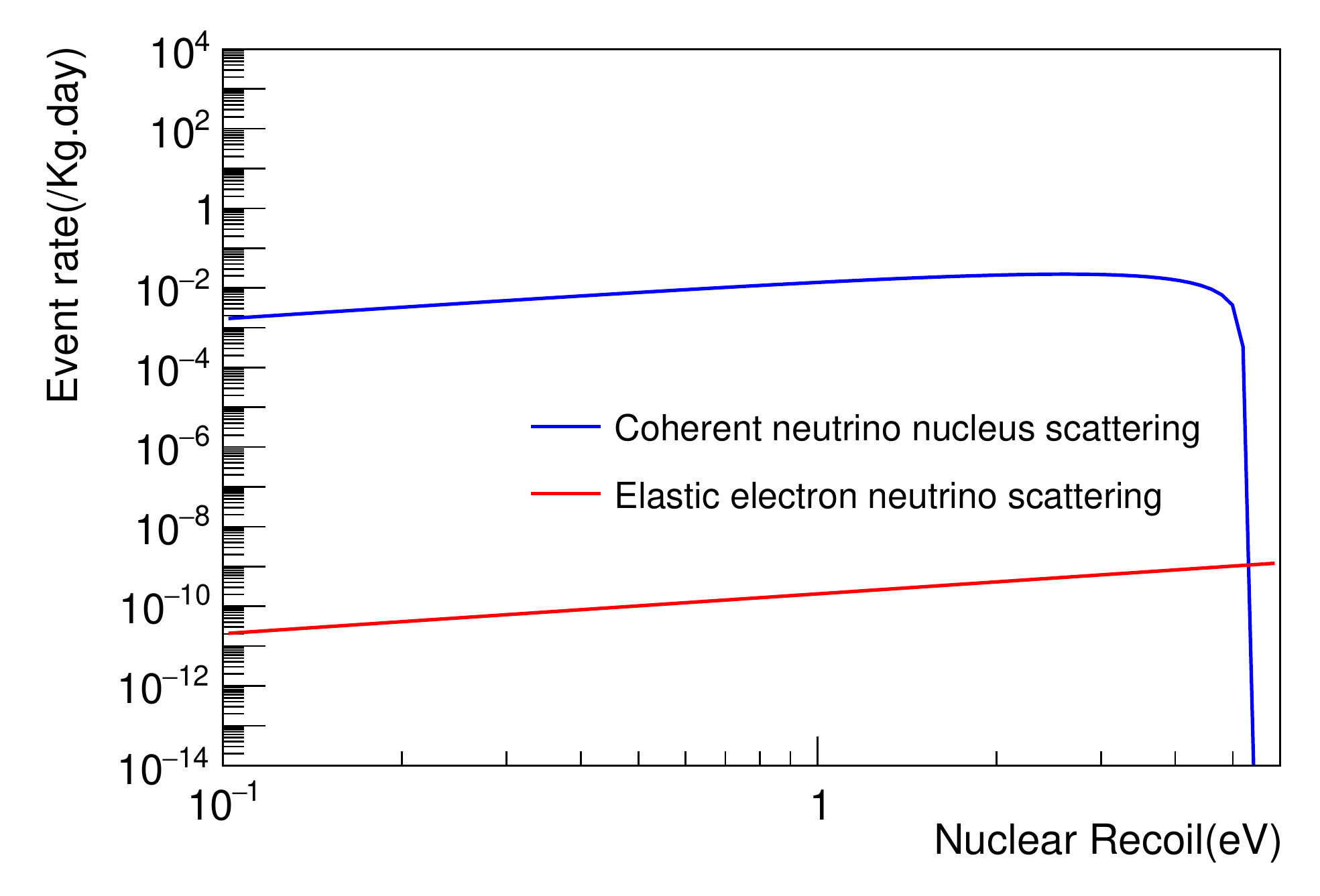}
    \caption{Event rate for pp solar neutrinos in a Ge detector of different threshold energies for elastic neutrino-electron scattering and coherent neutrino-nucleus scattering.}
    \label{fig:my_label6}
\end{figure}
\begin{figure}
    
     \includegraphics[width=0.5\textwidth]{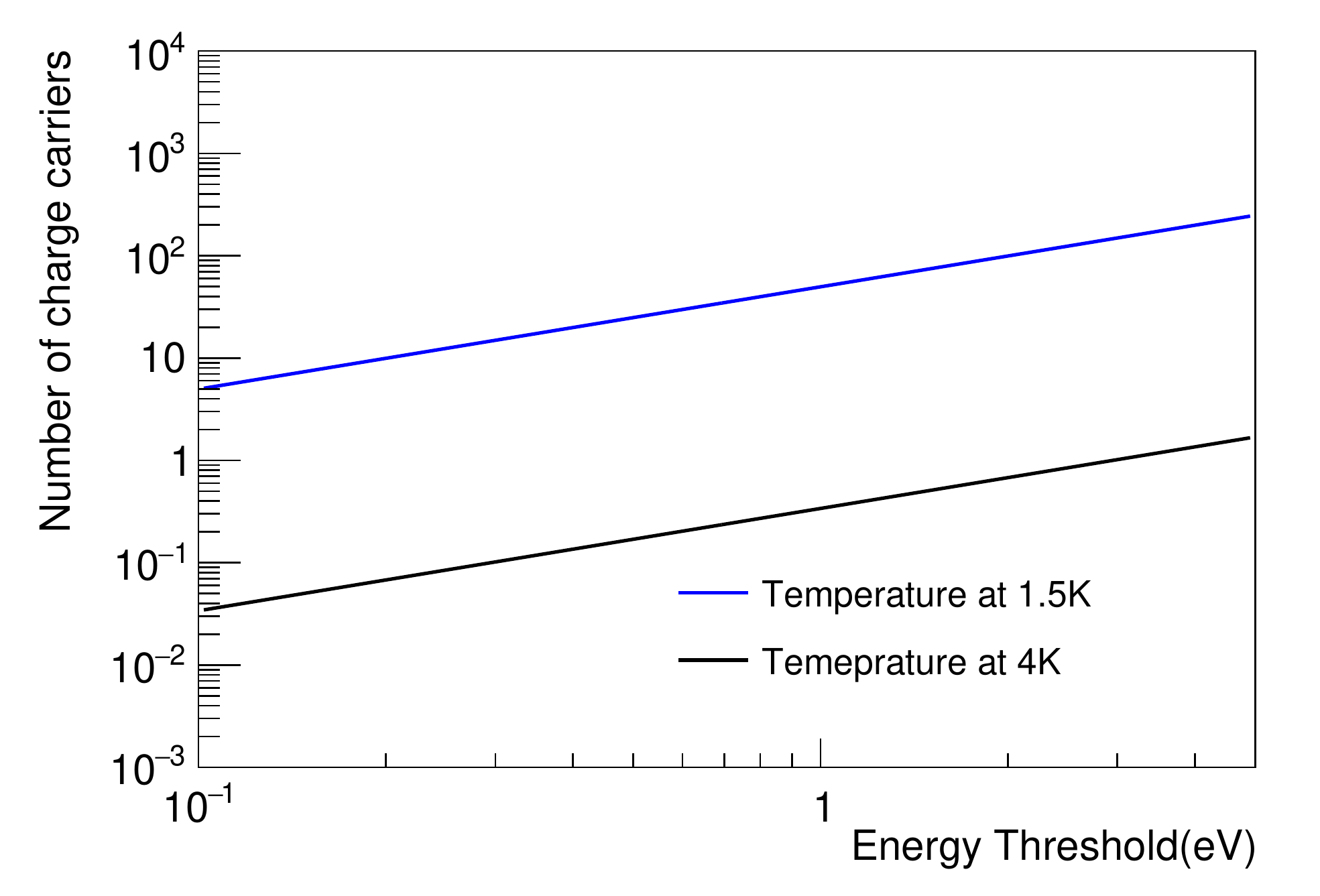}
    \caption{Number of charge carriers produced by phonons produced by pp neutrinos in a Ge detector for different threshold energies.}
    \label{fig:my_label7}
\end{figure}
\begin{figure}
    
     \includegraphics[width=0.5\textwidth]{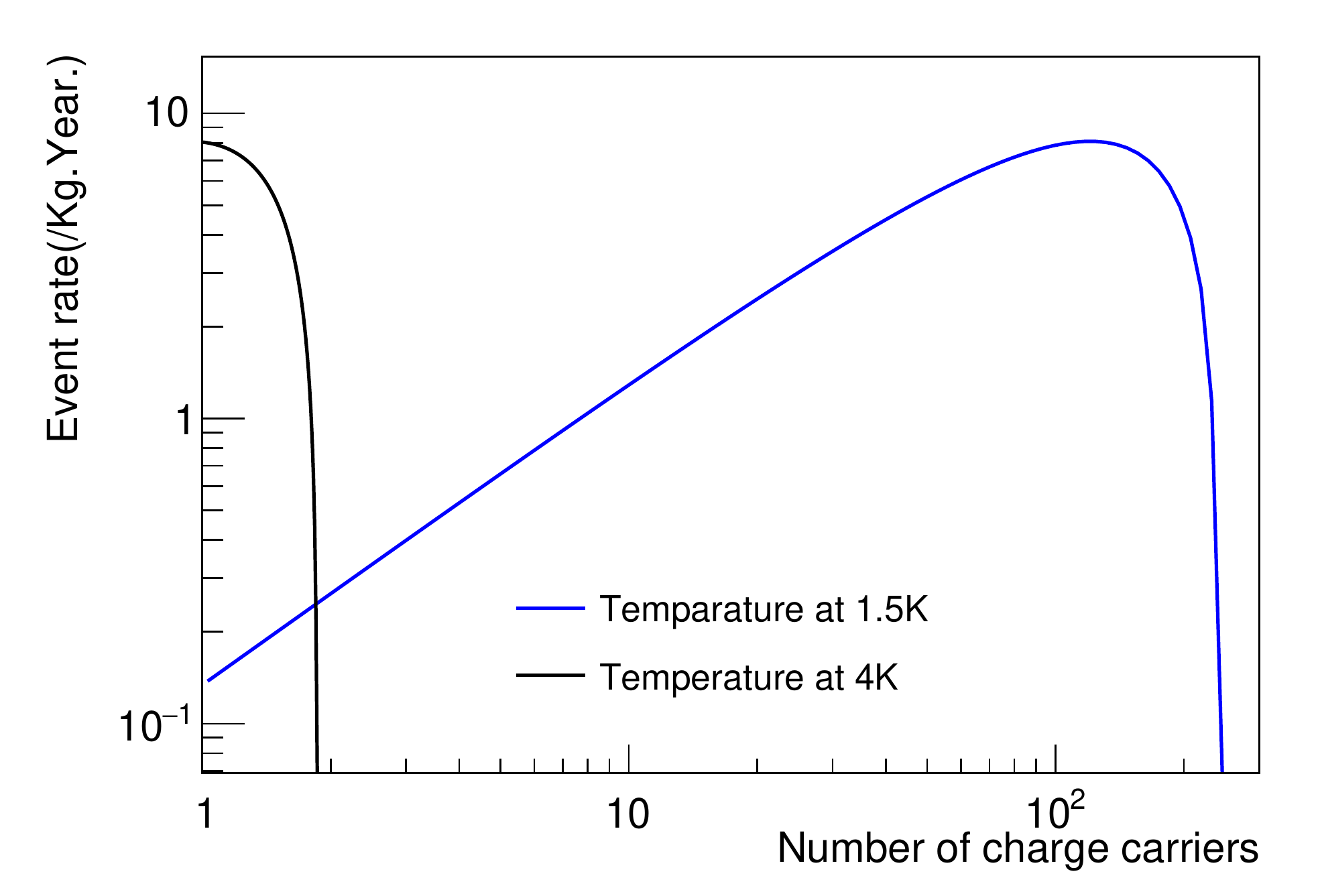}
    \caption{Event rate of pp neutrinos in a Ge detector when detector is operated at 1.5 K and 4 K.}
    \label{fig:my_label8}
\end{figure}

\begin{figure}
    \centering
    \includegraphics[width=0.5\textwidth]{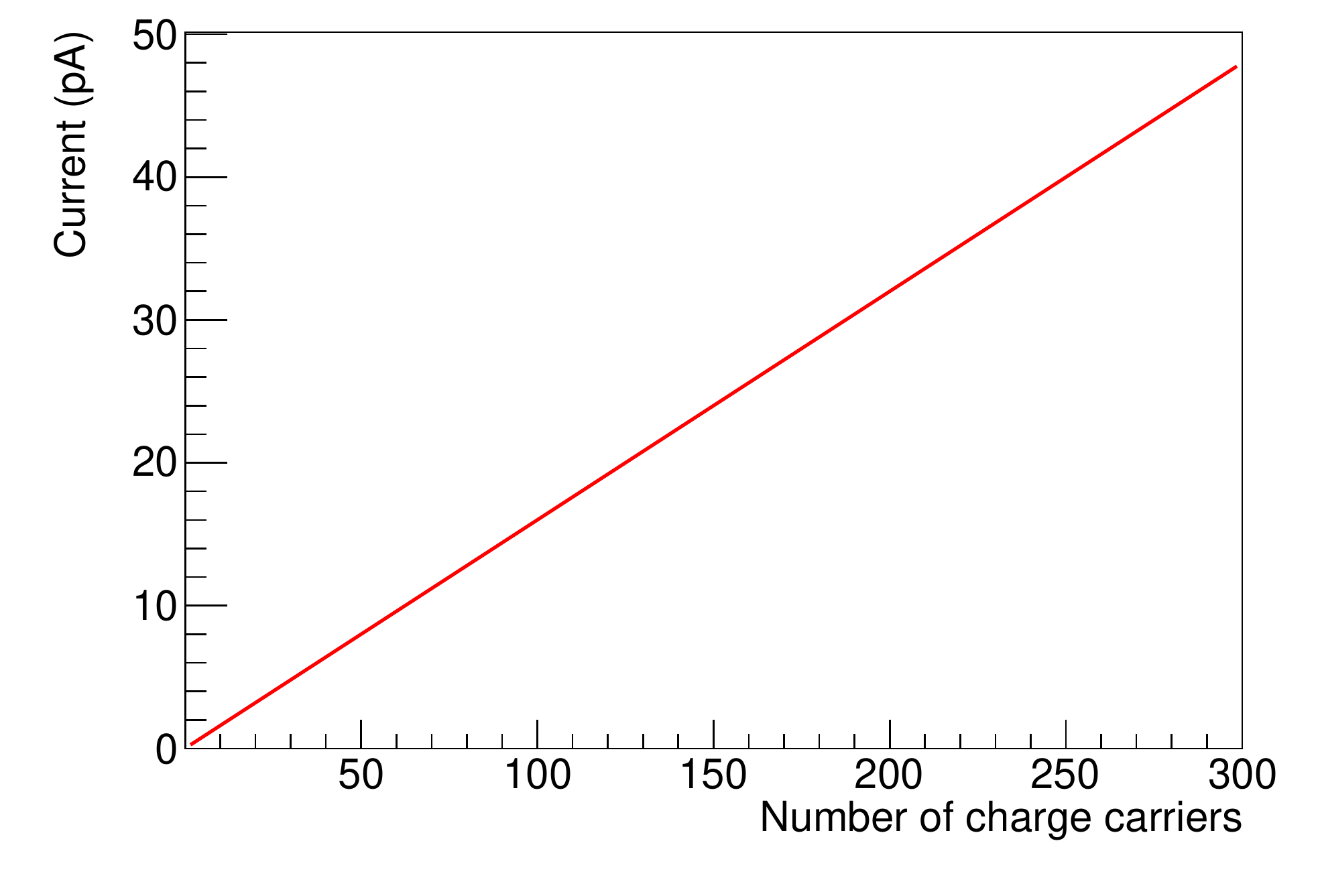}
    \caption{Current signal in the Ge detector when the pp neutrinos produce different number of charge carries through phonon channel}
    \label{fig:my_label9}
\end{figure}
At 1.5 K temperature, the total number of charge carriers generated in a detector with net impurity $3\times10^{10}/cm^3$ is about 90 as shown in Figure~\ref{fig:my_label5}. Note that these charge carriers are created by the different generations of the daughter phonons, from the  parent phonons of 0.026 eV energy. With an impurity level of $7\times10^{10}/cm^3$, at least one charge carrier can be produced when the detector is operated at 4 K.\\

Utilizing the flux and energy of $pp$ neutrinos~\cite{papoulias2018novel} in  equation ~\ref{equation_1} and ~\ref{equation_2} , we plotted the energy threshold versus the event rate for coherent neutrino-nucleus scattering (the blue curve) as shown in Figure ~\ref{fig:my_label6}. Note that when neutrinos interact with electron via exchange of neutral $z_0$ bosons, the differential cross-section is given by
\begin{multline}
\label{equation_7}
\frac{d\sigma_{ES}}{dT_r}=\frac{G_f^2m_e}{2\pi}[(g_\nu+g_a)^2+ (g_\nu-g_a)^2(1-\frac{T_r}{E_\nu})^2+\\(g_a^2-g_\nu^2)\frac{m_eT_r}{E_\nu^2}]
 \end{multline}
 where $m_e$ is the electron mass, $T_r$ is the electronic recoil, $E_\nu$ is the energy of incoming neutrinos, $g_v$ and $g_a$ are the vectorial and axial coupling respectively, and are deﬁned such that 
\begin{equation}
    g_\nu = 2sin\theta_w-\frac{1}{2},             g_a= -\frac{1}{2}
\end{equation} 
where $sin^2\theta_w$ is equal to 0.223~\cite{billard2015solar}. Using the flux and energy of $pp$ neutrinos~\cite{papoulias2018novel} in  equation ~\ref{equation_2} and ~\ref{equation_7}, the event rate for neutrino-electron scattering versus recoil energy is shown as the red curve in Figure ~\ref{fig:my_label6}. It is clear that the event rate induced by neutrino-electron scattering (the red curve) is much smaller than that of coherent neutrino-nucleus scattering (the blue curve). Therefore, coherent neutrino-nucleus scattering dominates in the low-energy region, which is studied in this work.

The event rate for $pp$ neutrinos is maximized at 2.6 eV  for 1 kg exposure for a year. Figure ~\ref{fig:my_label7} shows the variation of the number of charge carriers for different detection threshold energies. We can see that at two extremely low temperatures 1.5 K and 4 K, the number of charge carriers increases as energy increases for $pp$ solar neutrinos in a Ge detector. For example, the event rate is maximum at 2.6 eV where the number of charge carriers is $\sim$110 at 1.5 K and $\sim$0.5 at 4 K. The event rate variation versus the number of charge carriers is shown in Figure \ref{fig:my_label8}. The detector is most sensitive around 2.6 eV where the number of charge carriers is $\sim$110 at 1.5 K. One can expect that a 10 kg detector with an exposure of one year would obtain $\sim$1000 events, which is a $\sim$3\% precision in terms of measuring the $pp$ neutrino flux.    

The number of charge carriers can be converted into electric current $I$ as
\begin{equation}
    I=n_{carriers}q/t
\end{equation}
where $n_{carriers}$ is the number of the charge carriers , $q$ is the unit of charge equal to $1.6\times10^-{19}$ coulombs and $t$ is the charge collection time in the detector. If one assumes $t$ = 1 micro second, we can project the amount of current when $pp$ solar neutrinos hit our detector as shown in Figure ~\ref{fig:my_label9}. If the detector is capable of amplifying charge carriers by a factor of 100 through internal charge amplification,  the value of the current obtained from a single charge carrier is in the order of pico amperes. Hence, this current can be collected by a detector as described by D-M Mei et al~\cite{mei2018direct} which is the principle of this work. To create a single charge carrier, a phonon with energy of 0.01 eV can excite or ionize the impurity atoms in a Ge detector. This means that the detector  threshold can be as low as 0.01 eV.

Note that the background events can come from external and internal sources. Since the proposed detector is to have a threshold of 0.01 eV and the region of interest (ROI) for detecting $pp$ neutrinos is between 0.1 eV to 5.2 eV, in such a low-energy window, we expected both external and internal background events are negligible. This is because: (1) the external radioactive backgrounds and the muon-induced backgrounds can be minimized when the detector is operated underground with a well shielded experimental setup~\cite{mei2018direct}; and (2) the radioactivity inside the detector is often to generate background events through the Compton (inelastic) scattering process. These inelastic scattering of electrons by $\gamma$ rays are usually in the energy region of keV, much larger than the ROI for eV-scale Ge experiments~\cite{alexander2016dark,kane1992inelastic}. A only notable and unavoidable background is due to elastic scattering of external neutrons originated from  $(\alpha, n)$ reactions. These neutrons can be avoided effectively by using appropriate shielding~\cite{agnese2017projected,armengaud2013background}. More detailed discussion about the backgrounds on this type of detector is discussed by Dongming et.al.~\cite{mei2018direct}. A main source of the background events is the correlated background from other solar neutrinos. Other than the above sources of background events, for a low threshold detector, a common source of background is the various sources of noise associated with the cooling system, electronics and cables. However, this can be usually resolved by using a good cooling system, better electronics and cables. Therefore, we assume this can be put in control in this paper. 

It is worth mentioning that there are several sources of systematic uncertainties in the ROI for detecting $pp$ neutrinos. The first is the systematic uncertainty due to the subtraction of background events in the ROI. For example, the prominent sources of the background are the other solar neutrinos such as $^8B$ and $^7Be$ etc. The values of neutrino fluxes for all solar neutrinos used in the evaluation of the systematic uncertainty are calculated by using high metallicity SSM~\cite{papoulias2018novel,villante2020relevance}. The $pp$ and $pep$ neutrino fluxes are determined with $\leq1\%$ accuracy. However,the uncertainty in the fluxes of other solar neutrinos varies from $6\%$ to $30\%$~\cite{villante2020relevance}. These large uncertainties in the flux is one of the main sources of systematic uncertainty. Table 1 shows the calculated total event rates for $pp$ neutrinos is 16.58 events/kg.year and the sum of event rate of all other solar neutrinos is 9.201 events/kg.year, which are spread over in a large energy range as stated in Table 1. In the ROI, a total of $\sim$1.2 events/kg.year from other neutrinos are expected. Although the uncertainty of the flux can be as large as 30\%,  the contribution to the background reduction in the ROI is much smaller than the expected signal events. Hence, the systematic uncertainty from  the background deduction is small. The second source of the systematic uncertainty is the detection efficiency of a single charge carrier. The proposed detector is able to detect a single charge carrier. Due to the complexity of charge trapping, the proposed detector may lose charges, which results in a limited charge collection efficiency. However, this uncertainty should be minimized under a high field in which the charge trapping is negligible.  The final uncertainty is associated with the amplification factor that amplifies a single charge carrier through internal charge amplification. The amplification factor (K) is given by $K=2^{h/l}$ where $h$ is the length of avalanche region and $l$ is the free electron path of inelastic scattering. The approximate value of $l$ and $h$ in a planar Ge detector of 3 cm thickness  at 4 K is about $0.5\SI{}{\micro\metre}$ and $5\SI{}{\micro\metre}$ respectively. This leads to an amplification factor of about 1000~\cite{mei2018direct}. However, the value of $K$ is governed by the electric field, the concentration of impurities,  and the gradient of temperature in the detector. Due to the uncertainties in these parameters, it is difficult to obtain a constant value of $K$. The spread of the $K$ value is likely to impact the stability of the detection threshold and hence causes the systematic uncertainty in detecting $pp$ neutrinos. In summary, there will be some system uncertainties in detecting $pp$ neutrinos using the proposed low-threshold detector. When designing a detector system, those systematic uncertainties should be minimized to be less than than the statistical error.

\section{\label{sec:level5}{Conclusion}}
We present a viable detection method for studying $pp$ neutrinos using coherent elastic neutrino-nucleus scattering in a novel Ge detector with internal charge amplification. The very low energy deposition of $pp$ neutrinos interacting with Ge nucleus is dissipated through the emission of phonons. The diffusion of those phonons will undergo anharmonic decay. It is the propagation of those phonons that will excite and ionize impurities in Ge, which will allow us to detect the energy deposition from $pp$ neutrinos as low as 0.01 eV. If a Ge detector can internally amplify the charge signal by a factor of 100, then the charge carriers of $\sim$100 can be detected with current of $\sim$1 pA, which is a normal signal from a Ge detector. 

\section{Acknowledgement}
  The authors would like to thank Christina Keller for a careful reading of this manuscript. This work was supported by NSF OISE-1743790, DOE FG02-10ER46709,  DE-SC0004768, the Office of Research at the University of South Dakota and a research center supported by the State of South Dakota.


\bibliographystyle{spphys}       
\bibliography{sanjay.bib}   

%
%

\end{document}